\def\year{2022}\relax
\documentclass[letterpaper]{article} 
\usepackage{aaai22}  
\usepackage{times}  
\usepackage{helvet}  
\usepackage{courier}  
\usepackage[hyphens]{url}  
\usepackage{graphicx} 
\usepackage{booktabs}
\usepackage{amsmath}
\usepackage{multirow}
\usepackage[table]{xcolor}

\newcommand{\cor}[1]{
\cellcolor{orange!40}
}

\DeclareMathOperator*{\argmax}{arg\,max}

\usepackage{natbib}  
\usepackage{caption} 
\DeclareCaptionStyle{ruled}{labelfont=normalfont,labelsep=colon,strut=off} 
\usepackage{algorithm}
\usepackage{algorithmic}

\usepackage{newfloat}
\usepackage{listings}
\floatstyle{ruled}
\newfloat{listing}{tb}{lst}{}
\floatname{listing}{Listing}
\title{Controlling Perceived Emotion in Symbolic Music Generation \\ with Monte Carlo Tree Search}
\author {
    Lucas N. Ferreira\textsuperscript{\rm 1},
    Lili Mou\textsuperscript{\rm 1},
    Jim Whitehead\textsuperscript{\rm 2},
    Levi H. S.  Lelis\textsuperscript{\rm 1}
}
\affiliations {
    \textsuperscript{\rm 1}Alberta Machine Intelligence Institute (Amii), University of Alberta\\
    \textsuperscript{\rm 2}Computational Media Department, University of California, Santa Cruz\\
    lnferrei@ualberta.ca, lmou@ualberta.ca, ejw@soe.ucsc.edu, levi.lelis@ualberta.ca
}

\usepackage{bibentry}

\begin{document}

\maketitle

\begin{abstract}
    This paper presents a new approach for controlling emotion in symbolic music generation with Monte Carlo Tree Search. We use Monte Carlo Tree Search as a decoding mechanism to steer the probability distribution learned by a language model towards a given emotion. At every step of the decoding process, we use Predictor Upper Confidence for Trees (PUCT) to search for sequences that maximize the average values of emotion and quality as given by an emotion classifier and a discriminator, respectively. We use a language model as PUCT's policy and a combination of the emotion classifier and the discriminator as its value function. To decode the next token in a piece of music, we sample from the distribution of node visits created during the search. We evaluate the quality of the generated samples with respect to human-composed pieces using a set of objective metrics computed directly from the generated samples. We also perform a user study to evaluate how human subjects perceive the generated samples' quality and emotion. We compare PUCT against Stochastic Bi-Objective Beam Search (SBBS) and Conditional Sampling (CS). Results suggest that PUCT outperforms SBBS and CS in almost all metrics of music quality and emotion.
\end{abstract}

\section{Introduction}
\label{sec:introduction}

Neural language models (LMs) are currently one of the leading generative models for algorithmic music composition \cite{yang2019deep}. Neural LMs are trained to predict the next musical token with a data set of symbolic music pieces \cite{huang2018music}. A major problem with neural LMs is the lack of control for specific musical features on the decoded pieces. For example, one cannot control an LM trained on classical piano pieces to compose a tense piece for a scene of a thriller movie. 
It is hard to control the generative process of these models because they typically have a large number of parameters, and it is not clear what parameters affect what musical features. 
Controlling the perceived emotion of generated music is a central problem in Affective Music Composition \cite{williams2015investigating}, with applications in games \cite{williams2015dynamic}, stories \cite{davis2014generating}, and sonification \cite{Chen2015}. 


Controlling neural LMs to generate music with a target emotion 
started to be explored only recently. Two prominent decoding methods are Conditional Sampling \cite[CS,][]{hung2021emopia} and Stochastic Bi-Objective Beam Search \cite[SBBS,][]{ferreira2020computer}. The former consists of sampling from the LM with conditional signals that represent a target emotion. The latter is a variant of beam search that steers the probability distribution of LMs toward a target emotion by using a music emotion classifier at decoding time. 
In this paper, we propose a new decoding algorithm inspired by AlphaZero \cite{silver2017mastering}. AlphaZero uses Predictor Upper Confidence for Trees \cite[PUCT,][]{rosin2011pucb} as a policy improvement operator. We use PUCT to change an initial distribution given by an LM to a distribution that has higher musical quality and conveys a target emotion in the decoded pieces. 

We start with a neural LM that represents a prior probability distribution over tokens. At decoding time, we run multiple PUCT iterations to build a distribution of node visits that improves the LM distribution and steers it towards a target emotion. This is achieved by using both a music emotion classifier and a music discriminator as the value function of PUCT, while the LM is used as its policy. The policy and value function jointly define the set of nodes evaluated in the search. The frequency of node visits offers a distribution that produces music with higher quality than the LM initial distribution, shifted to convey the emotion given by the emotion classifier. 
We sample the next token in the sequence from this new distribution.

We train a neural LM with a Linear Transformer \cite {katharopoulos2020transformers} on the unlabelled pieces of the VGMIDI data set \cite{ferreira_2019}. The VGMIDI data set is a collection of 928 piano pieces from video game soundtracks, where 200 pieces are labelled according to the circumplex model of emotion \cite{russell1980circumplex}. We extend the number of unlabelled pieces of the VGMIDI data set from 728 to 3,850. Following the approach of \citet{ferreira2020computer}, we train the music emotion classifier with the 200 labelled pieces of the VGMIDI data set by fine-tuning the Linear Transformer LM with a classification head.
We train the music discriminator with the same fine-tuning approach. This discriminator distinguishes real data from the data created by sampling from the LM.

We conducted two experiments to evaluate our PUCT decoding method. The first one evaluates the quality and emotion of the generated samples with respect to human compositions based on a set of objective metrics computed from the musical structure of the pieces. The second is a listening test to measure human subjects' perception of the generated pieces' quality and emotion. We compare our method to SBBS, CS, and human compositions. Results showed that PUCT outperforms both SBBS and CS in almost all objective metrics, generating pieces closer to human compositions when compared with the other methods. According to our listening test, PUCT has better emotion controllability and overall music quality than SBBS and CS.

The main contribution of this paper is a decoding algorithm based on PUCT to improve music quality and control the emotions of music generated with neural LMs. Another contribution is a larger data set of symbolic video game pieces for training neural LMs for music. To the best of our knowledge, this is the first work to use PUCT as an algorithm to decode language models. Moreover, this is the first time PUCT is used to control the perceived emotion in generated symbolic music. Given that PUCT is agnostic to the classifier used to steer the distribution of the language model, we believe our method can be used to control other musical features that can be discriminated by a classifier, such as genre \cite{Ferraro_2018}, composer style \cite{daniel_yang_2021}, and difficulty \cite{Ghatas_2022}. We also believe that PUCT can be used to decode and control language models in other sequential domains such as text.

\section{Related Work}
\label{sec:related_work}

Our work is related to the control of perceived emotion in the context of LMs, to expert systems that use rules to control for the perceived emotion of generated music, and to interactive evolutionary algorithms, which evolve music pieces with a perceived target emotion using human subjects to evaluate the fitness of the candidate solutions. We review the methods from each of these areas. 

\subsection{Controlling Language Models}

Controlling the emotion of musical LMs started to be explored only very recently, but researchers have already proposed several methods that operate at either training time or decoding time. This paper is mostly related to the latter category. 
SBBS \cite{ferreira2020computer} is a decoding-time algorithm that uses Stochastic Beam Search to guide the LM towards a given emotion by multiplying the probabilities of the LM with the probabilities of an emotion classifier. At every decoding step, SBBS samples the next beam from this resulting distribution. SBBS applies Top-k filtering when expanding the search space in order to control the quality of the generated pieces. 
Our method differs from SBBS because we do not directly combine the distributions of the LM and the emotion classifier, but we search with the guidance of both models to create a distribution of node visits, from which the next token is sampled. 

Conditional Sampling \cite[CS,][]{hung2021emopia} works both at training and decoding time. It consists of sampling from the LM with conditional signals that represent a target emotion. The LM has to be trained with labelled pieces that encode emotion control tokens as part of the pieces. \citet{zhao2019emotional} proposed a similar approach, where an emotion conditional signal is added to the input of the LM to control the emotions of piano pieces. Our method is different than these previous approaches because it does not consider emotion as part of the language modelling task. Instead, we keep the original LM and define an auxiliary emotion classification task, using the emotion classifier to steer the original LM distribution while decoding the model. 

\citet{ferreira_2019} proposed a genetic algorithm to optimize the weights of an LM towards a given sentiment at training time. SentiMozart \cite{madhok2018sentimozart} is another example of training-time algorithm, which generates music that matches the emotion of facial expressions with independent LMs trained for different emotions. 
Our method differs from these training-time methods because we control the emotion of the pieces generated by the LM at decoding time.


\subsection{Expert Systems}

Expert systems are one of the most common methods for Affective Music Composition \cite{williams2015investigating}. They use rules based on music theory to control the perceived emotion of generated pieces. 
For example, \citet{williams2015dynamic} transform pre-generated melodies with an expert system to compose soundtracks that match the emotion of annotated video game scenes. TransPose \cite{davis2014generating} is a system that follows a similar approach to generate piano melodies for novels. MetaCompose \cite{scirea2017affective} is a framework designed to create background music for games in real-time, which generates a piece with a genetic algorithm and then uses an expert system to transform the final composition to match a target emotion. Our work is different than expert systems because we learn mappings from musical features to emotion directly from data.

\subsection{Interactive Evolutionary Algorithms}

Interactive Evolutionary Algorithms (IGAs) are another important class of methods for Affective Music Composition. IGAs evolve a random population of music pieces by evaluating them with human subjects, who judge whether or not the candidate pieces match a target emotion. For example, \citet{kim2004composing} proposed an IGA to compose polyrhythms for four percussion instruments. \citet{zhu2008emotional} presented an IGA based on the KTH rule system \cite{friberg2006overview} to create affective performances of pre-composed music pieces. \citet{nomura2018music} designed a distributed IGA to generate four-bar piano melodies with controllable brightness. Our method is different from IGAs because IGAs only provide a set of solutions at the end of the evolutionary process. Every time one wants to generate a new set of pieces, the slow interactive process must be repeated. Our method can quickly generate as many music pieces as needed.




\section{Language Model}

We use the linear transformer proposed by \citet{katharopoulos2020transformers} as a music LM, since it is one of the state-of-the-art methods for symbolic music generation \cite{hsiao2021compound}. We train this LM with the unlabelled pieces of the VGMIDI data set. Originally, the VGMIDI data set had 728 unlabelled pieces, but we expand it to 3,850 pieces in this work. The new pieces are piano arrangements of video game music created by the NinSheetMusic community\footnote{\url{https://www.ninsheetmusic.org/}}. The average duration and tempo of all $3,850$ pieces are $128.84$ seconds and $137.50$ beats per minute, respectively. We use the VGMIDI data, as opposed to EMOPIA \cite{hung2021emopia} or other data sets of symbolic music (e.g., MAESTRO), because with this increased number of unlabelled pieces, VGMIDI allows us to pre-train a relatively large LM and fine-tune it for emotion classification using pieces of the same genre.

There are multiple encoding methods for training music LMs but none of them have became a standard so far: piano roll \cite{dong2018musegan}, MIDI-like \cite{oore2017learning}, REMI \cite{huang2020pop}, or CP \cite{hsiao2021compound}. 
We encoded the VGMIDI pieces using REMI because it is mutually close to the encodings used in our baselines SBBS \cite{ferreira2020computer} and CS \cite{hung2021emopia}, making it easier to compare these different methods. REMI is a beat-based representation that has a vocabulary size of 338 tokens. It uses a bar and sub-beat tokens to represent time information. A bar token represents the beginning of a new bar, and a sub-beat token points to one of the constant numbers of sub-beat divisions in a bar. A note is represented by three independent tokens: a pitch token, a duration token, and a velocity token. Additionally, REMI uses a tempo token to control the pace of the music. The original REMI encoding uses a chord token to represent the chord of a particular point of the piece. We do not use that token, so as to keep it consistent with the original SBBS encoding~\cite{ferreira2020computer}. 

\section{Emotion Classifier and Discriminator}\label{sec:emotion_classifier}

Next we describe the emotion classifier and the discriminator we use to guide the generation of music with emotion.

\subsection{Emotion Classifier}
\cite{ferreira_2019} and \cite{ferreira2020computer} showed that
fine-tuning an LM with an extra classification head yields a
better model than training a classifier from scratch with the
same architecture of the LM. We follow a similar approach by fine-tuning a Linear Transformer instead of
an LSTM \cite{ferreira_2019} or a GPT-2 \cite{ferreira2020computer}.
We fine-tune this Linear Transformer with the labelled pieces of the VGMIDI data set.

In this paper, we define symbolic music emotion classification as a multi-class problem. We consider four different ``emotions'':
high valence and arousal \textbf{(E1)}, low valence and high arousal \textbf{(E2)},
low valence and arousal \textbf{(E3)}, and high valence and low arousal \textbf{(E4)}.
Each labelled piece in the VGMIDI data set has a valence label $v \in \{-1, 1\}$ and a arousal label $a \in \{-1, 1\}$. We map a pair of values to a categorical label (E1, E2, E3, or E4) by getting the quadrant in which the tuple $(v,a)$ lies in. Thus, a piece with values $(1,1)$ is mapped
to E1, $(-1,1)$ is mapped to E2, $(-1,-1)$ is mapped to E3,
and $(1, -1)$ is mapped to E4. This mapping yielded 76 pieces with label E1, 38 with label E2, 27 with label E3, and 59 with E4.

We use this model of emotion based on quadrants because the VGMIDI data set does not have enough data to cover a wider range of basic emotions relevant for music (e.g., fear, anger, joy, sadness, and surprise). Moreover, this is the same model used by CS \cite{hung2021emopia}. Alternatively, SBBS frames symbolic music emotion classification as two independent binary problems, one for valence and one for arousal \cite{ferreira2020computer}.
The multi-class approach simplifies the search task of controlling emotion of the generated pieces by having a single emotion model instead of two.

Since our search evaluates the emotion of sequences of varied length (see next section), 
we train our emotion classifier with
prefixes of varied lengths from the labelled pieces. These prefixes are created by splitting a given piece at every bar token of the piece. During training, the classifier
learns to map a varying-length sequence to an emotion. 
%
We denote our emotion classifier by $E$ and write $E(s, e)$ to denote the probability of a sequence $s$ being perceived of conveying emotion $e$.

\subsection{Music Discriminator}

We train a Music Discriminator to evaluate the quality of music pieces during search, guiding MCTS towards a distribution of node
visits that maximizes music quality jointly with the probability of the generated piece being perceived as of a target emotion. This discriminator works exactly like in a Generative Adversarial Network \cite{goodfellow2014generative}: it distinguishes real data from the data created by sampling from the LM. The architecture of our discriminator is similar to the music emotion classifier. We fine-tune the Linear Transformer LM with an extra classification head that outputs 1 for human-composed pieces and 0 for decoded pieces. 

We train the discriminator by combining the 200 labelled pieces of the VGMIDI data set, which were not used in the training process of the LM, with another 200 pieces generated with Top-p sampling \cite{holtzman2019curious} from the LM. The 200 labelled pieces are our real data and the sampled pieces are our ``fake'' data. The discriminator is trained to separate these two classes. During search, we use the output of discriminator, i.e., the probability of a piece being real, as a quality metric the pieces explored during search.

Since our search discriminates sequences of varied length (see next section), 
we also train our music discriminator with
prefixes of varied lengths from the real and fake pieces, following the same approach used to train the emotion classifier. This way the discriminator
learns to discriminate pieces of different lengths. 
%
We denote our music discriminator by $D$ and write $D(s)$ to denote the probability of a sequence $s$ being real.

\section{PUCT for Decoding Music Pieces with a Perceived Emotion}
\label{sec:puct}

Monte-Carlo Tree Search (MCTS) is a heuristic search algorithm traditionally used to
play board games with large search spaces \cite{browne2012survey}. Before taking
an action at the current game state, MCTS explores
the branches of the search tree, looking ahead to determine how different game moves could lead to stronger or weaker positions of the game. MCTS variations use different algorithms for deciding which branches of the tree to explore next. For example, UCT \cite{Kocsis2006} uses the UCB1 formula for deciding which branches to expand in each iteration of the algorithm, while AlphaZero uses the PUCT formula \cite{rosin2011pucb}.  

UCT uses only a value function to decide which branches to expand. AlphaZero, on the other hand, uses both a policy and a value function. AlphaZero showed that PUCT could improve trained policy networks in different games. That evidence inspired our hypothesis that PUCT could also improve a pre-trained neural music LM (our "policy" network) and steer its probability distribution towards a given emotion. Thus, similarly to AlphaZero, our approach also employs PUCT to generate music with a specific perceived emotion.

In our approach, PUCT receives REMI's starting token $s_0$; a LM $L$; an emotion classifier $E$; a music discriminator $D$, a maximum probability mass $p$ for Top-p sampling,
a search budget $d$ that defines the amount of search PUCT can perform before adding the next symbol to the musical sequence; and a target perceived emotion $e$. PUCT returns a sequence $s'$ starting with $s_0$ for which $E(s', e) \cdot D(s')$ is maximized. 

PUCT grows a search tree where each node $n$ in the tree represents a sequence of symbols from our vocabulary. The root of the tree represents the prefix $s_0$ provided as input. Each edge $(n, n')$ from node $n$ to node $n'$ represents the addition of a symbol to the sequence $n$ ($n'$ is one symbol longer than $n$) and we say that $n'$ is a child of $n$. Since each node $n$ represents a sequence $s$, we use $n$ and $s$ interchangeably.  
Initially, PUCT's tree is of size one as it contains the root of the tree representing $s_0$. In each iteration, PUCT performs the following steps to add a new node to the tree: (1) selection, (2) expansion, (3) simulation,  and (4) backpropagation. 

\subsection{Selection and Expansion}
In the selection step, for each node $n$ in the PUCT, we choose the symbol $l$ that maximizes the following equation:

\begin{equation}
\argmax_{l} Q(n, l) + c \times L(n, l) \times \frac{\sqrt{N(n)}}{1 + N(n, l)} \,,
\label{eq:puct}
\end{equation}

\noindent
where $Q(n, l)$ is a value of the emotion classifier combined with discriminator (explained in the simulation and backpropagation steps below), $c$ is an exploration constant, $L(n, l)$ is the LM's probability of adding symbol $l$ to the sequence $n$, $N(n)$ is the number of times node $n$ was visited in a selection step, and $N(n, l)$ is the number of times symbol $l$ was chosen at node $n$ in a selection step. Similarly to top-p, our implementation of the $\argmax$ operator in Equation~\ref{eq:puct} considers only the most likely symbols that sum up to a probability mass $p$. This 
allows the search to focus on sequences that are more promising according to the LM $L$. PUCT recursively selects nodes in the tree according to Equation \ref{eq:puct} until the symbol $l$ at a node $n$ leads to a sequence whose node $n'$ is not in the PUCT tree. In the expansion step, the node $n'$ returned in the selection step is added to the PUCT tree and its statistics are initialized: $N(n') = 1$, $N(n', l) = 0$ and $Q(n', l) = 0$ for all top $k$ symbols $l$ according to the probability $L(n', l)$.

\subsection{Simulation}

In the simulation step, we perform a roll-out from the recently added node $n'$. The roll-out consists of feeding the sequence $n'$ as input to the LM and auto-regressively sampling tokens from the LM until a bar token is found. We use Top-p sampling during the roll-out and denote the simulated sequence as $n_r'$. The roll-out isn't shorter than a bar because the classifier and the discriminator were trained on sequences with a minimum size of a bar. The roll-out isn't longer than a bar because it is unlikely that the resulting piece would reflect the piece's emotion decoded so far. At the end of the roll-out, we compute the $Q(n, l)$-value (recall that adding $l$ to $n$ generated the node $n'$) as follows:

\begin{equation*}
    Q(n, l) =
\begin{cases}
   E(n_r') \cdot D(n_r'), \, \text{if} \argmax(E(n_r')) = e \vspace{0.1cm} \\ 
   (1.0 - E(n_r', e)) \cdot (D(n_r') - 1.0), \text{otherwise}
\end{cases}
\end{equation*}
\vspace{0.1cm}

If the emotion classifier $E$ classifies the simulated sequence $n_r'$ as being of the target emotion $e$, then $Q(n, l)$ is non-negative as the product of two probability values ($E(n_r', e) \cdot D(n_r')$) is non-negative. Otherwise, $Q(n, l)$ is non-positive, which is given by $(1.0 - E(n_r', e)) \cdot (D(n_r') - 1.0)$. This conditional reward function is designed to simulate the result of a game where the score is positive if the target emotion is found, and negative otherwise. The probability $D(s')$ given by the discriminator is used as a penalty if the emotion of $s'$ is correct according to $E(s')$, and as a bonus if the emotion is wrong. After computing the $Q(n, l)$-value, $N(n, l)$ is set to $1$ as this is the first time $l$ is selected in node $n$.



\subsection{Backpropagation}
In the backpropagation step, the value of $Q(n, l)$ computed in the simulation step is used to update the $Q$-values of all the node-symbol pairs visited in the selection step. This is achieved by following the path in the tree chosen in the selection step in reverse order and updating the statistics of each node $n$ and node-symbol pairs $(n, l)$ as follows.

\begin{align*}
Q(n, l) &= \frac{Q(n, l) \cdot N(n, l) + E(n', e)}{N(n, l) + 1} \\
N(n, l) &= N(n, l) + 1 \,.
\end{align*}

The $Q(n, l)$-values are the average reward values of sequences with prefix given by $n$. 
The value of $Q(n, l)$ estimates how good continuations of the sequence $n$ added to the symbol $l$ are with respect to the discriminator and the emotion classifier. The backpropagation step completes an iteration of PUCT.

Equation \ref{eq:puct} ensures that sequences $n$ that maximize the value of $E(n, e) \cdot D(n)$ are visited more often in the PUCT iterations as they will have larger $Q$-values. Equation \ref{eq:puct} also accounts for the probability given by the language model, giving preference to sequences with higher probability according to $L$. The term $\frac{\sqrt{N(n)}}{1 + N(n, l)}$ in Equation \ref{eq:puct} certifies that all nodes have a chance of being explored by the search.

\subsection{Choosing the Next Token}
PUCT performs $d$ iterations before deciding which symbol is added to the sequence represented at the root of the tree. Let $n$ be the root of the tree. The symbol $l$ that will be added to the sequence $n$ is sampled from the distribution given by the values $\frac{N(n)}{\sum_{l} N(n, l)}$. The node $n'$ resulting from the addition of $l$ to $n$ becomes the new root of the tree and we perform another PUCT search with budget $d$ to choose the next symbol to be added to $n'$. This process is repeated until the next symbol added is the end-of-music symbol or a given number of tokens have been generated.

The PUCT search can be seen as an operator that improves the probability distribution over symbols given by the LM while accounting for the target emotion and discriminator. This is because the distribution given by $\frac{N(n)}{\sum_{l} N(n, l)}$ will favor symbols that lead to pieces that are more realistic, according to the LM and discriminator, while matching the target emotion as nodes representing such pieces are visited more often during search.

\section{Empirical Evaluation}

Based on the work of \cite{hung2021emopia}, we evaluate our PUCT approach according to the quality of the generated pieces and the method's accuracy in conveying a target emotion. We used objective and subjective metrics for both criteria. For objective quality evaluation, we used music structure metrics to evaluate whether the generated samples are similar to human compositions. We also computed the rate of generated pieces considered real by the discriminator. For objective evaluation of emotion, we used our emotion classifier as a proxy for human evaluators, computing the rate of the generated pieces that convey a target emotion. As a subjective evaluation of quality and emotion, we performed a listening test via Amazon Mechanical Turk (MTurk), asking human subjects about the quality and emotion they perceive in the generated pieces.
We compared PUCT with SBBS \cite{ferreira2020computer}, CS \cite{hung2021emopia}, and human-composed pieces from the VGMIDI data set. 

\subsection{Experimental Setting}

For PUCT, SBBS, and CS, we generated 20 music pieces for each emotion (E1, E2, E3, and E4), yielding $20\times4 = 80$ pieces per method. All pieces are 16 bars long and have $\frac{4}{4}$ time signature. For the human-composed pieces, we selected 20 random pieces with $\frac{4}{4}$ time signature from the VGMIDI data set for each emotion. We encoded all human-composed pieces using REMI and kept only the first 16 bars of each encoded piece.

We trained a Linear Transformer LM \cite {katharopoulos2020transformers} with $8$ transformer blocks and a maximum sequence length of 1,024 tokens. We used $8$ attention heads and an embedding layer of size 512. The size of the feed-forward layers in each transformer block was set to 1,024. We optimized the LM weights  with the Adam optimizer for $10$ epochs with mini-batches of size 16 and a learning rate of $1 \times 10^{-4}$. We saved the model at the end of every epoch and selected the one with the minimum validation loss. To simplify our music language modeling task, we used only the VGMIDI pieces with $\frac{4}{4}$ time signature to train the LM. This subset has 2,520 pieces, of which we used 2,142 (85\%) for training and 378 (15\%) for testing. All unlabelled pieces were augmented by (a) transposing to every key, (b) increasing and decreasing the tempo by 10\%, and (c) increasing and decreasing the velocity of all notes by 10\%, as \citet{oore2017learning} described. 

We trained the emotion classifier by fine-tuning our Linear Transformer LM with an extra classification head. The emotion classifier was trained with the 200 labeled pieces of the VGMIDI data set. We used 140 (70\%) pieces for training and 60 (30\%) for testing. We also trained the discriminator by fine-tuning our Linear Transformer LM with an extra classification head. The discriminator was trained with 400 pieces, the 200 labeled pieces (real) of the VGMIDI data set, and the other 200 (fake) pieces generated with the trained Linear Transformer LM via Top-p sampling ($p=0.9$). The emotion classifier and the discriminator were trained independently with the Adam optimizer for 100 epochs with mini-batches of size 16 and a learning rate of $1 \times 10^{-5}$. We saved these models at the end of every epoch and selected the ones which maximized accuracy on the testing set.

After training, the cross-entropy losses of the Linear Transformer LM are 0.32 (training set) and 0.54 (validation set). The emotion classifier and discriminator accuracies are 64\% and 93\% on their respective testing sets. At decoding time, we set $p = 0.9$ to filter tokens in the LM distribution for PUCT, SBBS, and CS. For SBBS, the size of the beam $b$ is set to $5$ and $k$ (for Top-k filtering) to $10$. For PUCT, we set the search budget $d$ (number of simulation steps) to $50$ and the exploration constant $c$ to $1$. It is important to highlight that, during search, both PUCT and SBBS have the same budget of $50$ evaluations to the emotion model.


\subsection{Objective Evaluation}

Following \citet{hung2021emopia}, we use three metrics to evaluate whether the generated samples are similar to human compositions: pitch range (PR), number of unique pitch classes used (NPC), and number of notes being played concurrently (POLY). These are music structure metrics computed with MusPy \cite{dong2020muspy}. Table \ref{tab:objective} reports the average results for these metrics and the rates of generated samples having the target emotion according to the emotion classifier (E) and classified as real by the discriminator (D). For PR, NPC, and POLY, the closer to the values of the human pieces, the better. For the emotion classifier and discriminator rates, the higher the value, the better. We report the results per emotion to evaluate how well each method performs in different target emotions. 

Overall, PUCT is better than all the other methods for all metrics and emotions. Among the music structure metrics, PUCT is not the best only in the POLY metric for emotion E1 and in the NPC metric for emotion E4, where CS is better than PUCT. This suggests that when compared to pieces generated by SBBS and CS, PUCT’s compositions have a pitch structure closer to human-composed pieces for all emotions except E4. PUCT’s compositions are also closer in polyphony to human pieces for all emotions except E1. 

According to the emotion classifier, PUCT is better than SBBS and CS in conveying all target emotions. It is surprising that, according to the emotion classifier, our method outperforms the human-composed pieces for both emotions E1 and E4. This is evidence that our method effectively optimizes for the emotion classifier. 

We conjecture that the low rates for E3 are because E3 is the least represented class in the VGMIDI data set. The lack of labeled data for E3 might not be enough for the classifier to learn musical patterns that distinguish E3 from the other emotion classes. We hypothesize that increasing the amount of E3 training pieces will increase the classification accuracy for that emotion.

\begin{table}[!t]
\centering
\small
\setlength{\tabcolsep}{7.5pt}
\begin{tabular}{clrrrr}
                        \toprule
                        \multicolumn{2}{c}{} & \multicolumn{4}{c}{Emotion} \\ \cmidrule{3-6}
\multicolumn{1}{l}{}   &          & \multicolumn{1}{c}{E1} & \multicolumn{1}{c}{E2} & \multicolumn{1}{c}{E3} & \multicolumn{1}{c}{E4} \\ \midrule
\multirow{5}{*}{\textbf{Human}} & PR  & $41.00$ &  $40.00$ & $33.35$ & $44.85$  \\
                       & NPC      &  $9.95$ &  $8.70$     &   $9.45$ &   $9.50$ \\
                       & POLY     &  $2.51$ &  $2.83$     &   $2.74$ &   $2.83$ \\ \cmidrule{2-6}
                       & E  &  $55\%$          &     $40\%$           &    $25\%$     &   $47\%$      \\
                       & D &  $68\%$          &     $73\%$           &    $78\%$   &  $72\%$     \\ \midrule
\multirow{5}{*}{\textbf{PUCT}}  & PR  &   $\mathbf{34.70}$ &  $\mathbf{40.75}$ &  $\mathbf{33.45}$  & $\mathbf{ 37.25}$ \\
                       & NPC      &    $\mathbf{7.35}$  & $\mathbf{8.00}$    &   $\mathbf{8.30}$    &    $7.75$  \\
                       & POLY     &   $2.01$         &  $\mathbf{ 3.00}$   &  $\mathbf{3.00}$     &    $\mathbf{2.66}$  \\ \cmidrule{2-6}
                       & E  &     $\mathbf{71\%}$          &   $\mathbf{34\%}$           &   $\mathbf{25\%}$             &       $\mathbf{55\%}$               \\
                       & D &    $\mathbf{58\%}$           &        $ 50\%$      &   $44\%$                   &         $\mathbf{52\%}$             \\ \midrule
\multirow{5}{*}{\textbf{SBBS}}  & PR  &    $31.40$  &   $28.80$  &  $31.75$    &    $30.45$    \\
                       & NPC      &    $5.45$  &   $5.25$   &  $6.55$     &   $6.50$    \\
                       & POLY     &    $1.97$ &    $1.75$  & $2.08$      &    $2.21$  \\ \cmidrule{2-6}
                       & E  &    $59\%$          &            $30\%$            &    $7\%$    &    $32\%$ \\
                       & D &    $24\%$          &           $28\%$             &      $31\%$  &    $29\%$                   \\ \midrule
\multirow{5}{*}{\textbf{CS}}    & PR  &   $30.40$  &   $33.60$  &    $33.10$  &  $34.60$  \\ 
                       & NPC      &     $6.70$  &   $7.55$  &    $7.10$     &   $\mathbf{8.80}$  \\
                       & POLY     &     $\mathbf{2.37}$   &   $2.67$ &   $2.22$      &   $2.44$  \\ \cmidrule{2-6}
                       & E &        $35\%$                &      $18\%$  &     $6\%$            &   $39\%$            \\
                       & D &       $48\%$                 &      $\mathbf{57\%}$     &   $\mathbf{51\%}$        &    $50\%$            \\
                       \bottomrule
\end{tabular}
\caption{Objective comparison between PUCT, SBBS, and CS with respect to human-composed pieces. For Pitch Range (PR), Number of Pitch Classes (NPC), and Polyphony (POLY), the values are averages, and the closer to the human-composed pieces, the better. For the Emotion classifier (E) and Discriminator (D) metrics, the values are rates, and the higher, the better. A value in bold is the best value of a metric (row) for a particular emotion (column). The values of the human composition are not highlighted because these values are used as a reference to evaluate the decoding algorithms.}
    \label{tab:objective}
\end{table}

\subsection{Listening Test}

We conducted our listening test as a between-subject study where a participant is asked to listen to 4 music pieces from a given system, each with a different emotion. Inspired by the work of \citet{hung2021emopia}, we asked the participants to rate each piece on a five-point Likert scale with respect to 1) Valence: is the piece negative or positive; 2) Arousal: is it low or high in arousal; 3) Humanness: how well it sounds like a piece composed by a human; 4) Richness: is the content interesting; and, 5) Overall musical quality. As part of questions 1) and 2), we provided a figure of the Circumplex model of emotion and asked the participants to explain their scores in 1-3 sentences based on the figure. We have $20$ pieces per emotion for each method, so we conducted $20$ listening tests per method, where a test is a sequence of 4 pieces from the same method, each with a different emotion. This yielded a total of $20 \times 4 = 80$ independent tests. We recruited 1600 participants (i.e., turkers) on Amazon Mechanical Turk (MTurk), assigning 20 subjects per test. We paid \$1.5 USD for each turker to complete the experiment.

To reduce the noise caused by turkers responding randomly to maximize profit,
we included two extra pieces in the beginning of the experiment and one in the middle.
The first two pieces are presented to the turkers as practice tasks, where we ask them
to listen to a labelled human-composed piece from the VGMIDI data set. In this practice
task, we let the participants know that the piece is composed by a human and if their
valence and arousal scores match with the ground truth scores. The extra piece added in 
the middle of the experiment is randomly chosen from the two practice pieces. We repeat a 
practice piece to check the consistency of the participants. We filtered out
all participants that were not consistent in their answers. Among the consistent participants, we also filtered out the ones who gave exactly the same string to explain their emotion scores (questions 1 and 2) of the four evaluated pieces. 
This filtering process resulted in 199 valid answers.
The average age of the kept participants was $40.36 \pm 11.43$, and the average formal musical training was $3.06 \pm 1.26$, where 1 means none, 2 elementary school level, 3 high school level, 4 undergraduate level, and 5 post-graduate level. All participants were located in the United States.

Table \ref{tab:subjective_emotion} shows the average scores of the questions 1) 
Valence and 2) Arousal for each method with respect to human-composed pieces. We present the results for each target emotion condition: High Valence ($\uparrow$$v$), Low
Valence ($\downarrow$$v$), High Arousal ($\uparrow$$a$), and Low Arousal ($\downarrow$$a$). For each condition, we ran a Kruskal-Wallis test across methods and if
significant, a pairwise Wilcoxon Rank-sum test with the Holm
correction. The methods are significantly different (p-value $\leq$ $0.05$) across all conditions based on the Kruskal-Wallis Test. Our method is significantly better than SBBS and CS in the arousal dimension ($\uparrow$$a$ and $\downarrow$$a$). Surprisingly, our method is significantly better than the human compositions in low arousal ($\downarrow$$a$). Moreover, there is no statistical difference between our method and human compositions in high arousal ($\uparrow$$a$) and high valence ($\uparrow$$v$). There is no statistical significance between PUCT, SBBS, and CS on the valence dimension ($\uparrow$$v$ and $\downarrow$$v$).

\begin{table}[!t]
\centering
\small
\setlength{\tabcolsep}{2.4pt}
\begin{tabular}{rcccc} \toprule
                      & \textbf{Human} & \textbf{PUCT} & \textbf{SBBS} & \textbf{CS} \\ \midrule
$\uparrow v$ & $3.92^a \pm 1.08$  & $\mathbf{3.70^{ab} \pm 1.15}$   & $3.50^b \pm 1.19$ & $3.38^b \pm 1.19$   \\
$\downarrow v$  & $2.58^a \pm 1.19$  & $\mathbf{3.14^b \pm 1.30}$   & $3.55^b \pm 1.14$ &  $3.23^b \pm 1.30$  \\
$\uparrow a$ & $3.62^a \pm 1.14$  & $\mathbf{3.65^a \pm 1.15}$   & $3.04^b \pm 1.11$ & $3.05^b \pm 1.17$   \\
$\downarrow a$  & $3.08^a \pm 1.15$  & $\mathbf{2.26^b \pm 1.05}$   & $2.71^c \pm 1.22$ & $3.16^a \pm 1.24$   \\

\bottomrule
\end{tabular}
\caption{Average scores (in 1–5) of valence and arousal as given by the human subjects in our listening test. We present the results for each target emotion condition: High Valence ($\uparrow$$v$), Low
Valence ($\downarrow$$v$), High Arousal ($\uparrow$$a$), and Low Arousal ($\downarrow$$a$). For $\uparrow$$v$ and $\uparrow$$a$, the higher the score, the better. For $\downarrow$$v$ and $\downarrow$$a$, the lower the score, the better. Values with shared letter superscripts$^{abc}$ in a given row were not significantly (p-value $> 0.05$) different in pairwise post-hoc comparisons. A value in bold is the best score for a particular emotion (row). The scores of the human compositions are not highlighted because these scores are used as a reference to evaluate the decoding algorithms.}
\label{tab:subjective_emotion}
\end{table}

Table \ref{tab:subjective_quality} shows the average scores for questions 3) Humanness, 4) Richness, and 5) Overall Quality. Similarly to Table \ref{tab:subjective_quality}, for each condition, we ran a Kruskal-Wallis test across methods and, if significant, a pairwise Wilcoxon Rank-sum test with the Holm correction. The methods are significantly different (p-value $\leq$ $0.05$) across all conditions based on the Kruskal-Wallis Test.
PUCT is significantly better than CS in humanness ($H$) and significantly better than SBBS in richness ($R$). In terms of overall quality ($O$), our method significantly outperforms CS and SBBS.

These results show that PUCT achieves state-of-the-art performance in decoding music pieces with perceived emotion, improving arousal expression and overall music quality. Moreover, since PUCT and SBBS have the same search budget, we can also conclude that PUCT is more efficient in exploring the space of pieces to control the LM towards target emotions than SBBS. Moreover, Table \ref{tab:subjective_emotion} suggests that, in the generated pieces, Arousal is easier than Valence for human subjects to distinguish.  

\begin{table}[!h]
\centering
\small
\setlength{\tabcolsep}{2.5pt}
\begin{tabular}{ccccc} \toprule
                      & \textbf{Human}         & \textbf{PUCT}           & \textbf{SBBS}      & \textbf{CS} \\ \midrule
$H$     & $4.07^a \pm 0.79$    & $\mathbf{3.88^b \pm 0.84}$   & $3.73^{bc} \pm 1.02$ & $3.53^c \pm 1.03$   \\
$R$     & $4.11^a \pm 0.80$    & $\mathbf{3.82^b \pm 0.97}$        & $3.47^{c} \pm 1.10$   & $ 3.65^{bc} \pm 1.00$   \\
$O$      & $4.11^a \pm 0.73$    & $\mathbf{3.94^b \pm 0.76}$         & $3.62^c \pm 1.01$ & $3.68^c \pm 0.96$  \\
\bottomrule
\end{tabular}
\caption{Average scores (in 1–5) of Humanness ($H$), Richness ($R$), and Overall Quality ($O$) as given by the human subjects in our listening test. For each metric, the higher the score, the better. Values with shared letter superscripts$^{abc}$ in a given row were not significantly (p-value $>$ 0.05) different in pairwise post-hoc comparisons. A value in bold is the best score for a particular quality metric (row). The scores of the human compositions are not highlighted because these scores are used as a reference to evaluate the decoding algorithms.}
\label{tab:subjective_quality}
\end{table}



\section{Conclusion}

This paper introduced PUCT as a decoding algorithm for language models to generate musical pieces that convey a target emotion. We use the language model as PUCT's policy and a music emotion classifier combined with a music discriminator as PUCT's value function. The PUCT search improves the initial distribution given by the language model while steering it towards a target emotion. We evaluated PUCT with objective metrics based on the music structure of the generated pieces. Moreover, we conducted a listening test (i.e., user study) to evaluate how human subjects perceive the generated samples' quality and emotion. The results showed that PUCT could control the emotion of the generated pieces better than Conditional Sampling and SBBS. PUCT also improved the overall quality of the generated pieces with respect to these two methods. To the best of our knowledge, this is the first application of PUCT for decoding language models.

\section{Acknowledgements}

We would like to thank the NinSheetMusic community for their work on producing piano arrangements of video game music and for kindly helping us get the music pieces in MIDI format. 

\bibliography{aaai22}

\end{document}